\documentclass[preprint,showpacs,preprintnumbers,amsmath,amssymb]{revtex4}

\usepackage{graphicx}
\usepackage{subfigure}
\usepackage{amssymb}
\usepackage[dvips]{color}

\begin{document}
\title{Structural stability and magnetism of FeN from first principles}
\author{A.~Houari$^{a)}$, S.~F.~Matar$^{b)}$\footnote{Corresponding author (s.matar@u-bordeaux1.fr)},
M.A. Belkhir$^{a)}$ and M. Nakhl$^{(c)}$}
\address{$^{a)}$Laboratoire de Physique Th\'eorique, Departement de Physique, Universit\'e de Bejaia. Bejaia, Algeria. \\
$^{b)}$Institut de Chimie de la Mati\`ere Condens\'ee de Bordeaux, CNRS,  Universit\'e Bordeaux 1. Pessac, France.\\
$^{(c)}$ Facult\'e des Sciences et de G\'enie Informatique. Universit\'e St Esprit de Kaslik. Jounieh. Lebanon.}
\date{\today}

\begin{abstract}
In the framework of density functional theory (DFT), the structural and  magnetic properties of FeN mono nitride have been investigated using the all electrons augmented spherical wave method (ASW) with a generalized gradient GGA functional for treating the effects of exchange and correlation. Calculation of the energy versus volume in hypothetic rocksalt (RS), zinc-blende (ZB) and wurtzite (W) types structures shows that the RS-type structure is more stable than the others. Spin polarized calculation results at equilibrium volume indicate that the ground state of RS-FeN is ferromagnetic with a high moment, while ZB-FeN and W-FeN are non magnetic. The influence of distortions on the stability is taken into account by considering FeN in two different face-centred tetragonal structures (fct): fct-rocksalt and fct-zincblende. The magnetovolume effects with respect to Slater-Pauling-Friedel model are discussed. The electronic structures analyzed from site and spin projected density of states are reported. A discussion of the structural and magnetic properties of FeN is given with respect to N local environment of Fe. 
\end{abstract}
\maketitle

\section{Introduction}
The study of the iron-nitrogen system has attracted much scientific importance for basic research as well as for technology. It was extensively investigated over the two decades (1950-1970) following its first discovery by Jack ~\cite{Jack}. A large number of results have been reported for Fe-rich nitrides, see \cite{Matar} and references therein. This was motivated by their potential applications such as pigments for high density magnetic recording \cite{ieee}. On the contrary, only few reports are available in the case of the mononitride FeN~\cite{Heinman,ouled}, which is considered as interesting material in the emergent spintronic field (sources of spin injection to semiconductors or diluted magnetic semiconductors). Recently, thin films of FeN have been synthesized and two possible crystal structures were reported ~\cite{Nakag,Hino,suzuki,rissanen}. One is the rocksalt {\it (RS)} structure with a lattice constant of a = 8.638 bohr, and by performing $^{57}$Fe M\"ossbauer spectroscopy measurements the authors suggested rocksalt FeN to be an antiferromagnet~\cite{Nakag,Hino}. The other structure is of zinc-blende {\it (ZB)}-type with a lattice constant of a = 8.185 bohr   and a micromagnetic character~\cite{suzuki}. 

In order to achieve a better understanding of their physical properties and to discuss the experimental discrepancies, some theoretical investigations have been undertaken. From the available literature a clear controversy exists between the studies carried out in the framework of density functional theory. Shimizu et al. ~\cite{shimizu} have concluded that a ferromagnetic rocksalt FeN is more stable than non magnetic zinc-blende nitride. The same results were reported later on the stability of the FeN RS-type structure, but without a complete agreement for the magnetic ordering ~\cite{filip,Kong}. However, Lukashev et al.~\cite{lukash} have joined some others~\cite{eck} in predicting a stable non magnetic ZB structure for FeN. In this context and in order to provide an improved theoretical explanation, we carried out new comprehensive spin density functional calculations in both ZB and RS phases. Further, for a better investigation of the magnetic properties with respect to the structural aspects, we have also considered FeN in the hexagonal wurtzite-type structure. With the sophisticated techniques of synthesis reached recently (synthesis of thin films and nanowires, \ldots), the structural distortions may have important effects on the investigated physical properties. To take into account the effects of some possible distortions, we have also investigated FeN in the face centered tetragonal fct structure. As a matter of fact, the manganese mononitride (MnN) has been recently found in (fct-RS) -type structure i.e., a tetragonally distorted rocksalt structure~\cite{suzuki1}. In our case, we have considered FeN in both tetragonally distorted rocksalt structure (fct-RS) and tetragonally distorted zinc-blende (fct-ZB) structure for sake of completeness. To the best of our knowledge, this is the first theoretical study of FeN in wurtzite, fct-RS and fct-ZB type structures.
The structural, electronic and magnetic properties are thus reviewed by analyzing the calculated electronic structure, total energies and crystal field influence. A description of the computational details is given in section 2. Our results for the calculated total energy and magnetic moment are presented in section 3, with a discussion of the magnetovolume effects. Section 4 is devoted to the Slater-Pauling-Friedel model as applied to iron nitrides. In section 5 the non magnetic results are discussed in the framework of the Stoner theory of band ferromagnetism. Section 6 details the electronic structure via partial density of states analysis. Finally a conclusion and a prospective for future developments are given in the last section.

\section{Computational details}

Our first principles calculations are performed in the framework of the density functional theory ~\cite{kohn,sham} using the generalized gradient approximation (GGA) with the Perdew, Burke and Ernzerhof PBE parameterization for the exchange and correlation ~\cite{PBE}. This parameterization scheme was preferred over the LDA (local density approximation) in view of the results on iron nitrides cited in this work~\cite{shimizu,Kong}. Self-consistent calculations were carried out using the scalar-relativistic augmented spherical wave method (ASW) ~\cite{kubler,eyert} based on the atomic sphere approximation (ASA). In this method, the value of the GGA-PBE $kappa$  was fixed to 0.8. In the ASW method, the wave function is expanded in atom-centered augmented spherical waves, which are Hankel functions and numerical solutions of Schr\"odinger's equation, respectively outside and inside the so-called augmentation spheres. In order to optimize the basis set, additional augmented spherical waves were placed at carefully selected interstitial sites. The choice of these sites as well as the augmentation radii were automatically determined using the sphere-geometry optimization algorithm  \cite{sgo}. The basis set consisted of Fe(4s, 4p, 3d) and N(2s, 2p) valence states. The Brillouin zone integrations were performed with an increasing number of k-points (16x16x16) in order to ensure convergence of the results with respect to the k-space grid. The convergence criterion is fixed to 0.001 mRy in the self-consistent procedure and charge difference $\Delta~Q$=10$^{-4}$ between two successive iterations. 
In order to establish a reference for the spin-polarized calculations we started with a set of spin-degenerate calculations. We note that this does not represent a paramagnetic situation which would require heavy computations involving large supercells with random spin orientations. It allows assigning a role to the orbitals responsible of the magnetic instability toward spin polarization in a mean field analysis using the Stoner theory of band ferromagnetism~\cite{stoner}. 

\section{Results for the total energy and the magnetic moments}
In a collective electrons scheme, such as the one used here, the magnetization arises from interband spin polarization, i.e., it is mediated by the electron gas. This is opposed to the localized electron moments where magnetization arises from intraband spin polarization such as in oxide systems, especially insulating ones. From this, it is expected that  the magnetovolume effects will be large in intermetallic and insertion alloys systems. The theoretical equilibrium volume and the structural preference of the FeN compound are obtained by calculating the variation of the total energy versus the volume of the cubic cell in the rocksalt and zinc-blende structures. For wurtzite, {\it fct}-rocksalt and {\it fct}-zinc-blende, the same procedure (total energy versus the volume of the unit cell) was done twice, in order to obtain the two equilibrium lattice  constants a and c which determine the equilibrium volume.  The stability toward magnetism in each phase is given by comparing the spin-polarized (SP) and the non spin-polarized (NSP) total energy  values at theoretical equilibrum volume.  With the same k-points grid of the reciprocal lattice and for different pairs (E,V), after a fitting with a Birch equation of state~\cite{birch}, we obtain after convergence a quadratic curve with a minimum (E$_{0}$,V$_0$). Figures 1-a and 1-b show the energy versus volume curves of FeN
in the {\it RS} and {\it ZB} structures for NSP and SP cases, respectively. For the wurtzite structure, figure 1-c shows the energy variation versus volume at the equilibrium c lattice constant, i.e. the variation of energy
versus the a lattice constant. Because the {\it fct}-rocksalt  and {\it fct}-zinc-blende structures are derived from the cubic ones, the same variation is observed and the results are not plotted here. The results for theoretical lattice constants, total energy and other equilibrium properties are summarized in table I for all the studied structures with the experimental data for the lattice constants. \\
Figure 1-a shows that the ferromagnetic state of FeN in the RS phase is preferred at the calculated equilibrium volume to the non magnetic state. On the contrary, the magnetic order in ZB and W phases can be favored but only at higher volume, see figure 1-b and figure 1-c. At equilibrium, the energy difference between the ferromagnetic and non magnetic states in these two structures are nearly zero, thus instability toward magnetism could exist. This instability may be the reason of the micromagnetic character observed experimentally for FeN in ZB structure. From table I our estimated theoretical lattice constant in the SP case (a = 8.219 bohr) agrees well with the experimental data for the {\it ZB} phase. However, a large difference exists between the obtained theoretical lattice constant (a = 7.769 bohr) and the experimental reports (a$_{exp}$ = 8.638 bohr) in the {\it RS} structure. It is important to mention that our results for this point are in good agreement with all the theoretical studies of the FeN reported in the literature (even those predecting a {\it ZB} stable structure for FeN), but surprisingly, up to now, no one was able to reproduce the experimental data.\\
The common explanation put forward to explain this large difference is that it would originate firstly from surface effects because the samples were grown as thin films, and secondly from the nonstoichiometry of the elaborated samples. The values of the total energies  in table I indicate that the FeN compound in the
ferromagnetic state should be more stable in {\it RS} structure than in {\it ZB} or {\it W} ones. At equilibrium FeN in these two latter structures has a zero magnetic moment, while the unit cell of {\it RS}-FeN possesses a moment of $2.62 \mu_B$ (a moment of $2.51 \mu_B$ per Fe-atom). Our results are in a good agreement with those reported by Shimizu et al.~\cite{shimizu} concerning the ferromagnetic order of FeN. A value of  $2.67\mu_B$ was reported Y.Kong but in a stable antiferromagnetic (AFM) {\it RS}-FeN~\cite{Kong}. For W-FeN, we think that this is the first study of this hypothetic structure, and thus a comparison of our results with others' cannot be made. 

We also performed antiferromagnetic (AF) calculations at the ferromagnetic equilibrium lattice constant in order to check  for a possible instability toward such a magnetic configuration for the ground state as suggested by Nagakawa er al. ~\cite{Nakag} and Hinomura et al. ~\cite{Hino}. Among possible spin arrangements we examinded for this purpose the  [001]-alignement (AF-[001]) in which the spins are parallel within a layer but alternating direction between adjacents layers in [001] direction. From carefully converged calculations the results in Table I indicate that the energies of AF and ferromagnetic (FM) states for FeN in RS structure remain very close. Moreover, in {\it fct}-rocksalt structure, the former one seems more stable. The energy differences between AF and FM states are too small to make a clear conculsion on the nature of the magnetic order in the two crystal varieties. On the other hand, the energy difference between RS-FeN and {\it fct}-RS-FeN is very small, thus one can believe that under some elaboration conditions, FeN nitride can crystallize in a face-centred tetragonal-rocksalt structure. Because of this, the AF character observed experimentally can be explained either by this AF-FM instability in the {\it RS}-structure or by some lattice distortions which stabilize the FeN compound in AF-{\it fct}-rocksalt structure. \\ 
In earlier investigations~\cite{mohn} on the magnetovolume effects in the Fe$_{4}$N nitride, it has been shown that there is a transition from ``low moment-low volume'' to ``high moment-large volume'' which resembles the moment versus volume dependence of $\gamma$Fe. This behavior seems to occur in all structures studied here of equiatomic FeN nitride. As it can be seen in figure 2, the variation of the moment in {\it RS}-FeN (and in {\it fct-RS}-FeN which is not shown here) is very similar to those of Fe{\it II} (located at the faces of the cubic cell) in Fe$_{4}$N. In {\it ZB}-FeN (and in {\it fct-ZB}-FeN also not shown here) a sudden increase from zero to a value as high as $2.85\mu_B$ occurs at high volume, like the behavior of Fe{\it I} atoms (located in the corners) in Fe$_{4}$N. The same variation of the magnetic moment is indicated for {\it W}-FeN.\\
An analysis of the effects of the Fe-N spacing on the magnetic moment value shows that the Fe-N distances, corresponding to equilibrium of zinc-blende, {\it fct}-zinc-blende and wurtzite structures (see table I), give a non zero moment (around 1$\mu_B$) in rocksalt or/and {\it fct}-rocksalt structure. If we consider the similar increase of the magnetic moment value reported elsewhere in  CsCl-type FeN~\cite{Kong}, one should recognize that at different atomic environements (crystalline structures) even with the same Fe-N distance, different values of the moment are obtained. Thus the major effect which prevails in the value of the magnetic moment is the nature of the {\it d-p} hybridization in the Fe-N bond which is related to the atomic environment (the number of the nearst-neighbors), and the distance Fe-N has only an average effect.  

\section{Analysis within the Slater-Pauling-Friedel model} 
The magnetic moment variation with valence electron count can be discussed  by means of the Slater-Pauling (SP) model~\cite{stoner}. This is mainly illustrated by the SP curve which is a plot of the average magnetic moment $\mu_{av}$ with the valence electron count $Z_v$ for intermetallic alloy systems characterized by strong ferromagnetism. From the plot the variation of the average magnetization of an alloy with the solute concentration can be obtained. For most of the magnetic systems, either the majority ($\uparrow$) or minority ($\downarrow)$ spin population is known. The magnetization being provided by the difference of electron occupation between the majority ($\uparrow$) and the minoriry ($\downarrow$) spins, the following relationships can be established: m = n$\uparrow$ - n$\downarrow$  and Z$_{v}$ = n$\uparrow$ + n$\downarrow$. Hence m, the magnetic moment, can be obtained either as: $m  =  2~n(\uparrow)~ -~  Z_v$,  
if n($\uparrow$) is known, or m  = $Z_v~ - 2\,n(\downarrow )$, if n($\downarrow$) is known. These two expressions for m describe two branches with opposite 45$^{o}$ slopes of the Slater-Pauling  curve around which the experimental points are gathered. If one assumes that the magnetic moment mostly arises from d-band polarization, the quantity ``magnetic valence'' can be defined -here for systems where n$_{d~\uparrow}$ is known as $Z_m  =  2\,n_{d \hspace{+0.5mm}\uparrow} - \,\,  Z_v$.  Upon alloying, the electron count of the d states changes in a discrete way as treated by Friedel~\cite{friedel}, so that $n_{d\uparrow}$ is either 0 (early transition elements) or 5 (late transition elements). For instance, for Fe, Z$_{v}$ is 8 and Z$_{m}$= 10-8 = 2, whereas for N ($n_{d\uparrow}$ = 0), $Z_v=3$ and $Z_m= 0-3 = -3$. This leads to an alternative representation called Slater-Pauling-Friedel (SPF) curve where $\mu_{av}$ is plotted against $Z_m$: $\mu_{av}  =  \,\,  Z_m + \,0.6$. In this writing , the figure 0.6 is a nearly constant contribution to the magnetic moment arising from s and p electrons.  The calculated atom-averaged magnetic moments ($\mu_{av} $) via the expression above and from ASW calculations are summarized in Table II. The results of the present study (for FeN nitride) are shown in the last line of the table; the other data are from~\cite{matar1} in which the SPF model was applied for the whole series of the Fe-N systems. In their study, Matar and Mohn~\cite{matar1} have found a good agreement between SPF model and ASW calculation results, and all average moments decrease with increasing amount of nitrogen.  While our obtained moments in {\it ZB}-FeN, {\it fct-ZB}-FeN and {\it W}-FeN agree well with SPF model in finding a vanishingly small magnetization, a large discrepancy exists in the case of {\it RS}-FeN and {\it fct-RS}-FeN which carry large magnetizations. This interestingly points to the limits of the SPF model which is an average scheme dealing with the electronic configuration of the  chemical species in presence, and does not account for the nature of the crystalline structure and the chemical bonding. 
\section{Mean field analysis of non magnetic configurations} 
Within the Stoner theory of band ferromagnetism ~\cite{roy} which is a mean field approach, the large DOS at the Fermi level ${\it n(E_{F})}$ is related to the instability of the non magnetic state with respect to the onset of intraband spin polarization. When ${\it n(E_{F})I_{s} > 1}$ where $I_s$ is the Stoner integral, calculated and tabulated by Janak~\cite{janak} for the elemental systems, then the non magnetic state is unstable with respect to the onset of magnetization. In the {\it RS}-FeN type with a density of states at the Fermi level ${\it n(E_{F})} = 71.596$ states/Ry/Fe and a Stoner parameter $I_s = 0.034$ Ry the Stoner criterion is fulfilled (2.434). On the contrary, for the {\it ZB}-FeN and {\it W}-FeN we have  ${\it n(E_{F})} = 14.106$ states/Ry/Fe and ${\it n(E_{F})} = 15.152$ states/Ry/Fe respectively, the Stoner criterion (resp. 0.48 and 0.515) is not fulfilled. As a consequence, the magnetic behavior of FeN is consistent with the Stoner theory. As it was discussed in the literature, the magnetic moment of Fe atom in FeN ~\cite{shimizu,Kong,lukash} and generally in the Fe nitrides~\cite{Matar} is very sensitive to nearest-neighboring Fe-N distance and their d-p hybridization. Our results confirm that this latter is the most important effect. 

\section {Density of states in the ferromagnetic configuration}
Considering the results obtained for the electronic structure of FeN at equilibrium we discuss the electronic structure from the plots of the density of states (DOS). Spin polarization results from a nearly rigid band shift between the majority spins  ($\uparrow$) to lower energy and minority spin ($\downarrow$) to higher energy due to the gain of energy from exchange. The site and spin projected partial DOS (PDOS) are shown in fig. 3 for {\it RS}-FeN and {\it ZB}-FeN at the theoretical equilibrium lattice constants. The {\it W}-FeN PDOS which resemble those of {\it ZB}-FeN are not presented here. Energy low lying N(2s) states at $\sim$ -10 eV as well as ES(PDOS) of vanishingly low intensity are not shown. As a matter of fact ES receive charge residues from actual atomic spheres and they are known to ensure for covalency effects in such systems~\cite{Matar}. Note that in RS-FeN the octahedral environment splits the Fe{\it -3d} states into $t_{2g}$ and $e_{g}$ manifolds, whereas in the tetrahedral environment of {\it ZB}-FeN, the d states are split into $e$ and $t_2$ manifolds. These respective environments result in totally different DOS features within the two structures: whereas they are continuous over the VB in {\it RS}-FeN more localisation is observed for the {\it  ZB}-FeN. One consequence is that {\it RS}-FeN is a metal through the the itinerant $e_g$ states crossing the Fermi level while in {\it ZB}-FeN the system is close to an opening of gap at E$_{F}$. Within the valence band (VB)  two parts can be considered. While the lower energy part, centered at -5 eV, is mainly composed of N{\it 2p} states mixing with itinerant Fe{\it -$e_g$} ones, the higher energy part from -3 eV to 1 eV is dominated by the Fe{\it -$t_{2g}$} states which are localized and thus responsible of the magnetic moment ({\it RS}-FeN). These states are moved to lower energy for $\uparrow$ spin population (upper panel) and hence are found below the Fermi level E$_{F}$, whereas for the minority-spin ($\downarrow$) the DOS peaks are located above $E_F$ (lower DOS half panel). This is due to the exchange-splitting whose absence in the ZB-FeN PDOS causes mirror PDOS in upper and lower half panel of fig. 3b. Clearly the $T_d$ crystal field is unfavorable to the onset of magnetization in {\it  ZB}-FeN and {\it a fortiori} in {\it W}-FeN.

\section{Conclusion} 

With the use of the self-consistent DFT-based ASW method, we have investigated the structural, electronic and magnetic properties of the iron mononitride (FeN) in the rocksalt (RS), zinc-blende (ZB) and wurtzite (W) structures. We have also investigated the effects of distortion by considering FeN in a tetragonally distorted rocksalt fct-RS and in a tetragonally distorted zinc-blende fct-ZB structures. The calculated total-energy shows that FeN is stabilized in a ferromagnetic (FM)-rocksalt structure with a theoretical lattice constant a = 7.769 bohr ~and a magnetic moment of 2.62 $\mu_ B$. While ZB-FeN and W-FeN type seem to prefer a paramagnetic order at theoretical equilibrium, they possess nevertheless a nonzero magnetic moment at higher lattice constant. This involves strong magnetovolume effects which are equally accompanied by crystal field ones depending on whether the local environment is octahedral or tetrahedral.  The  equilibrium energies of non magnetic and FM states of ZB-FeN are very close and this instability toward magnetism may be the reason of the micromagnetic character observed experimentally. We have also found a very small energy difference between RS-FeN and fct-RS-FeN, thus one can predict that in some special synthesis conditions such as the presence of strains in the growth of layers of iron mononitride, FeN could crystallize in the latter structure. The energies of AF and FM states in RS-FeN and fct-RS-FeN are too close, so it is difficult to make a clear conclusion on the nature of the magnetic order of the ground state for these structures. As it was suggested in the literature, the present work confirms that the magnetic properties of FeN are dominated by $p-d$ hybridization of the Fe-N interaction. However, we have found that the Fe-N distance has only an average effect on the value of the magnetic moment. On the other hand, our results of FeN compound complete the finding on the competition between chemical and magnetovolume effects in the series of iron nitrides: Fe$_{8}$N, Fe$_{4}$N, Fe$_{3}$N as well as Fe$_{2}$N ~\cite{matar2}. For FeN in ZB and W-type structure a good agreement with Slater-Pauling-Friedel model, connecting average magnetization with valence, is found; but when RS-FeN is considered a large difference exists so that other effects related with the nitrogen environment of Fe (O$_{h}$ versus T$_{d}$ or C$_{4v}$) should become prevailing. 
As a prospective we think of completing the explanation of the FeN properties in two complementary directions: 
\begin{itemize}
\item by considering slab and surface calculations, and 
\item by taking into account non (i.e., sub- or super-) stoichiometries in extended lattices. This should comply with the nature of the samples elaborated experimentally which point to N substoichiometric FeN1 x compositions. 
\end{itemize}
Such investigations will call for heavy calculations involving other computational frameworks such as the 
use of pseudopotential based codes for geometry optimizations~\cite{vasp}. The studies are underway. 

\section{Acknowledgements}
One of us, SFM, acknowledges computational facilities provided by the M3PEC-Regional Mesocenter University Bordeaux 1 (Web site http://www.m3pec.u-bordeaux1.fr). Discussions at an early stage of this work with Prof. Dr Peter Mohn of the University of Vienna, Austria,  are equally acknowledged.

{}
\begin{table}[!tbp]
\caption{Calculated equilibrium properties (in atomic units(bohrs) 1Ry=13.6eV) for FeN nitride in 
five different structures: Rocksalt {\it (RS)} , zinc-blende {\it (ZB)}, wurtzite {\it (W)}, face centred tetragonal-rocksalt {\it (fct-RS)} and face centred tetragonal-zinc-blende{\it (fct-ZB)} }
\begin{ruledtabular}
\begin{tabular}{llllccc}
equilibrium& Lattice       & total      &$d_{Fe-N}(bohr)$&$m_{tot} (\mu_B)$\\
properties& constants (bohr)& energy (Ry)&               &                  \\
\hline
  NSP{\it -RS}   & a=7.610                 & -2655.026&3.788&\\
  SP{\it -RS}    & a=7.769              & -2655.047&3.788&2.60\\
        & a=7.559\footnote{Ref.  \cite{shimizu}}      &          &    &    \\
        & a=7.939\footnote{Ref.  \cite{Kong}}   &          &    &     \\
        & Exp: a=8.640\footnote{Ref.  \cite{Nakag}} &          &    &     \\
AF{\it -RS}   &                          & -2655.046  &3.890&2.36 \\
\hline
 NSP{\it -ZB}   & a=8.200                    & -2654.997&3.534& \\
  SP{\it -ZB}    & a=8.219                    & -2654.998&3.551 &0.00\\
        & a=8.24$^{a}$ & \\
        & a=7.933\footnote{Ref.  \cite{lukash}}  & \\
        & Exp: a=4.332\footnote{Ref.  \cite{suzuki}} & \\
\hline
 NSP{\it -W}   & a=5.778         & -2655.018&3.606 &\\
       & c=9.637      &           \\
 SP{\it -W}    & a=5.809                   & -2655.019&3.608&0.00\\
       & c=9.637                 &            \\
\hline
 NSP{\it -fct-RS}   & a=7.778        & -2655.022&3.869&\\
       & c=7.580      &           \\
 SP{\it -fct-RS}    & a=7.844                   & -2655.044&3.924&2.63\\
       & c=7.655                 &            \\
AF{\it -fct-RS}&      & -2655.045           &     3.924               &   2.48  \\
\hline
 NSP{\it -fct-ZB}   & a=8.090         & -2654.992&3.523\\
       & c=8.015      &           \\
 SP{\it -fct-ZB}    & a=8.128                   & -2654.993&6.663&0.00\\
       & c=8.015                 &            \\
\end{tabular}
\end{ruledtabular}
\end{table}

\begin{table}[!tbp]
\caption{Application of the Slater Pauming Friedel model of the magnetic valence to iron nitrides (see text).}
\begin{ruledtabular}
\begin{tabular}{llllccc}

Iron nitride & $Z_v$      & $Z_{m}$ & $m_{SPF}(\mu_B)$& $m_{ASW}(\mu_B)$\\
         &            &           &=$Z_m$+0.6       &\\
\hline
\hline
Fe$_{8}$N\footnote{Ref. \cite{matar1}}   & 7.44    &1.44  & 2.04&2.17  \\
Fe$_{4}$N$^{a}$  & 7.0 & 1.0&1.60&1.67\\
Fe$_{3}$N$^{a}$   & 6.75 & 0.75  & 1.35   &1.44    \\
Fe$_{2}$N$^{a}$    & 6.33 &  0.33 &  0.93  & 0.95    \\
FeN (from Fe$_{2}$N)$^{a}$   & 5.5 &  -0.5 &  0.10  & 0.00 \\
\hline
FeN & 5.5 & -0.5& 0.10  & 0.00 ({\it ZB})\\
                &       &   &       & 0.00 ({\it W})\\
                &       &   &       & 2.62 ({\it RS})\\
                &       &   &       & 2.60 ({\it fct-RS})\\
                &       &   &       & 0.00 ({\it fct-ZB}) \\
\end{tabular}
\end{ruledtabular}
\end{table}
\begin{figure}[h]
\begin{center}
\includegraphics[width=0.55\linewidth]{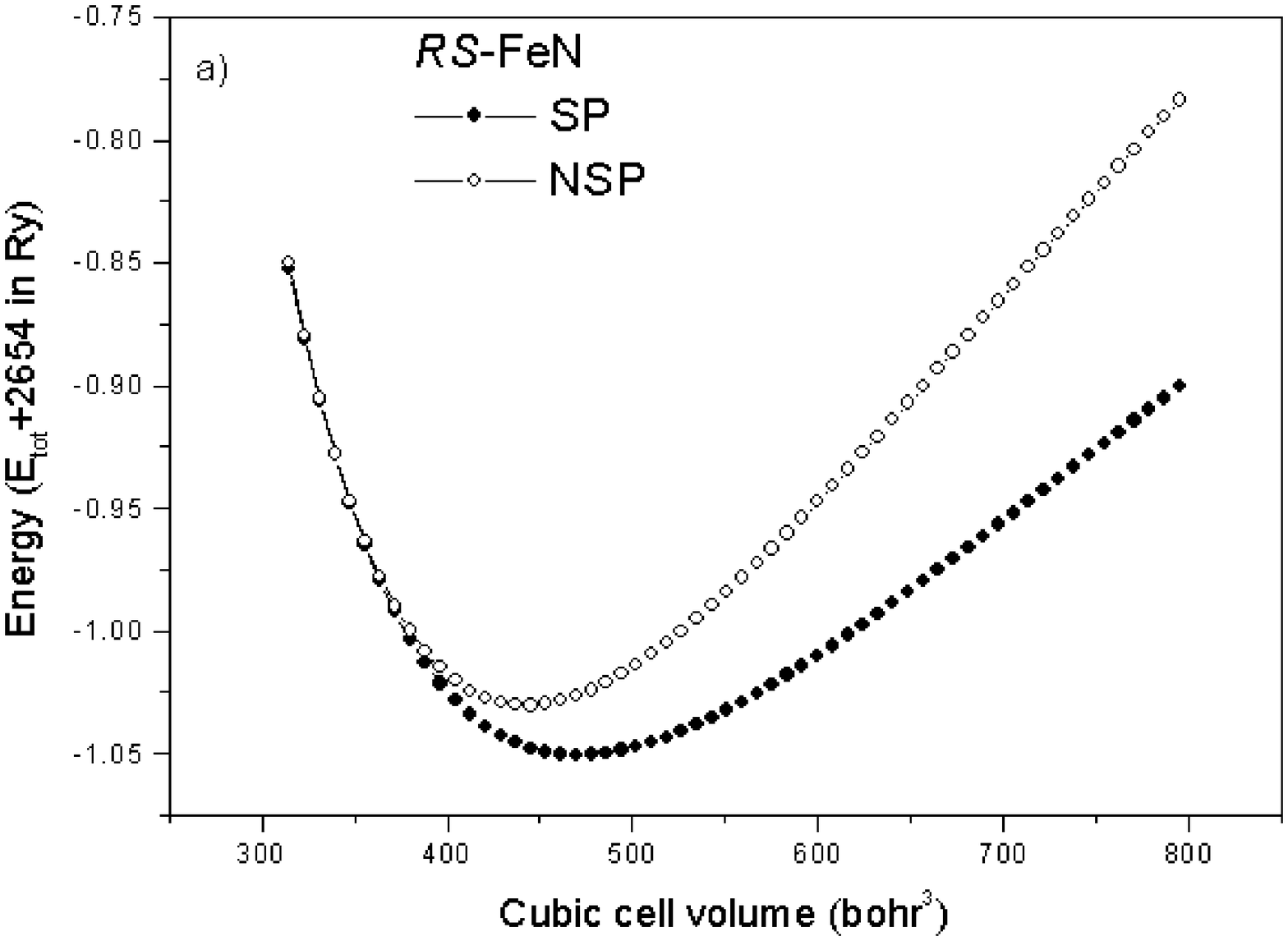}
\includegraphics[width=0.55\linewidth]{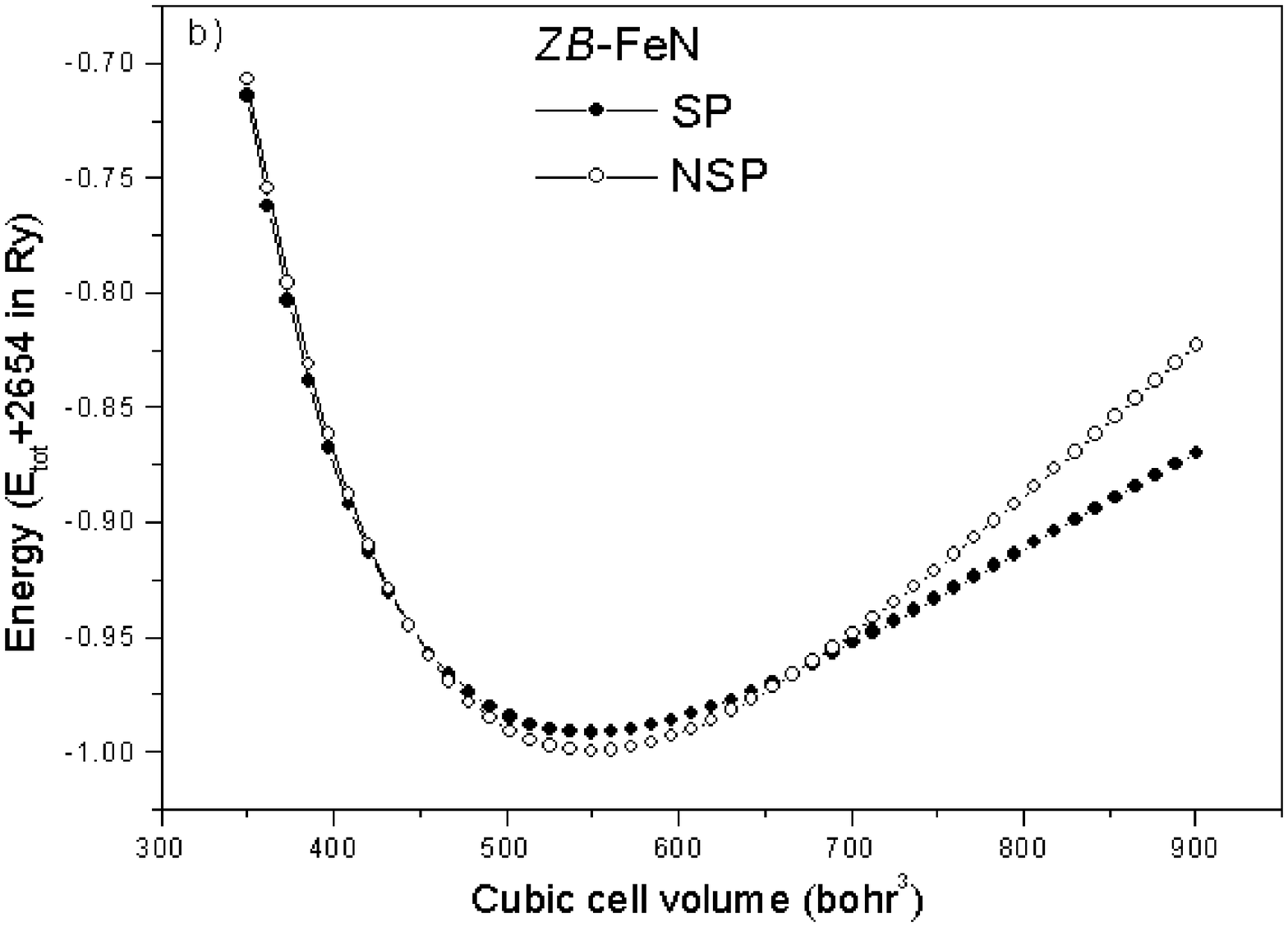}
\includegraphics[width=0.55\linewidth]{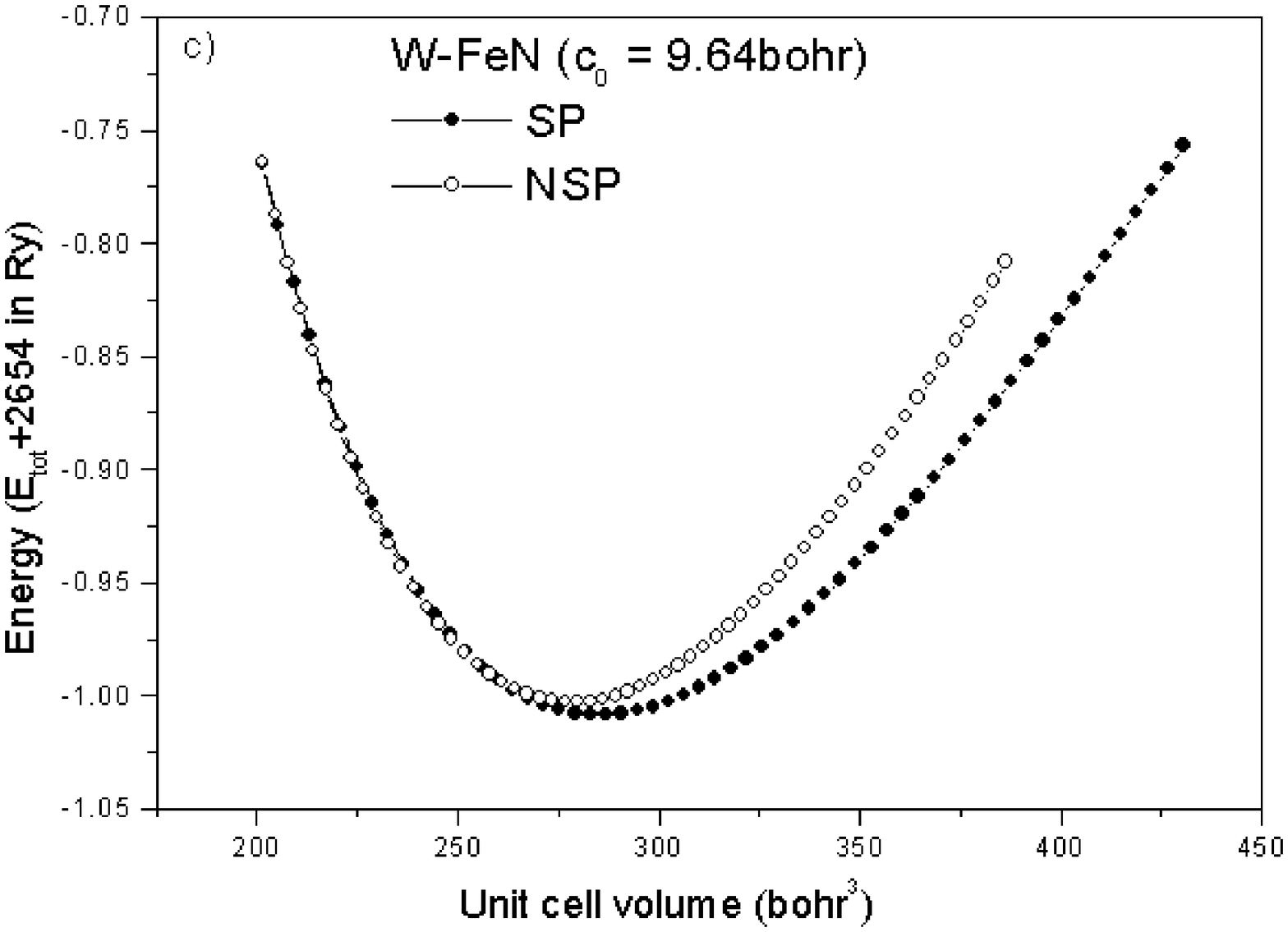}

\caption{Spin polarized (SP) and non-spin polarized ( NSP) total energy versus volume in: a) RS-FeN, b) ZB-FeN and c) W-FeN, at equilibrium c$_{0}$ lattice constant for the latter.}
\end{center}
\end{figure}
\begin{figure}[h]
\begin{center}
\includegraphics[width=10.5cm, height=9.5cm]{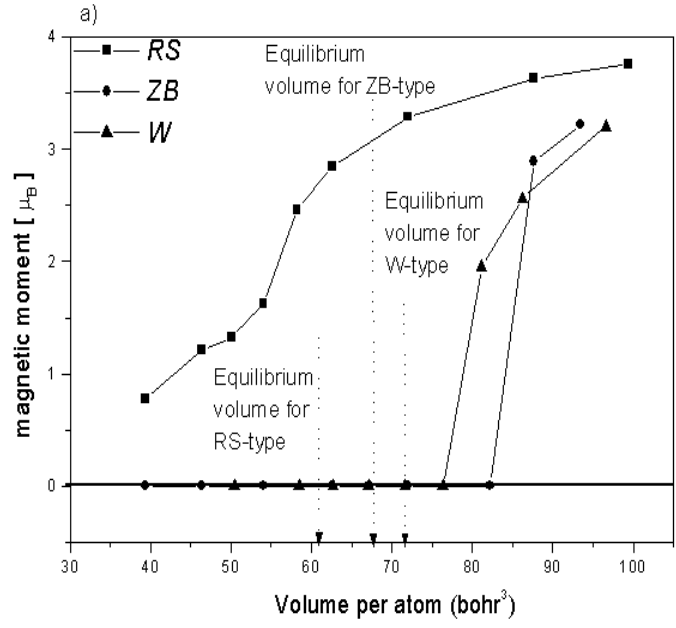}
\includegraphics[width=10.5cm, height=9.5cm]{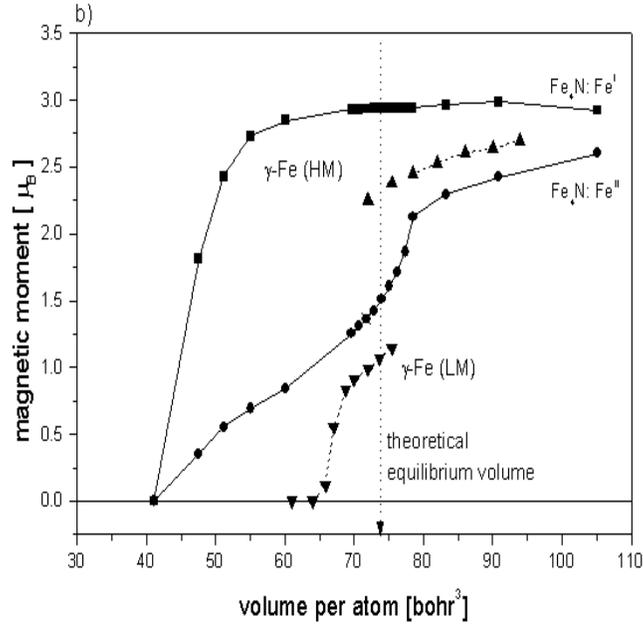}
\caption{ Magnetic moment versus volume in: a) present work: FeN in RS,  ZB and W structures. b) Fe$_{4}$N and $\gamma$ Fe (replotted from reference \cite{mohn}) }
\label{mfv}
\end{center}
\end{figure}
\begin{figure}[h]
\begin{center}
\includegraphics[width=9.5cm, height=8cm]{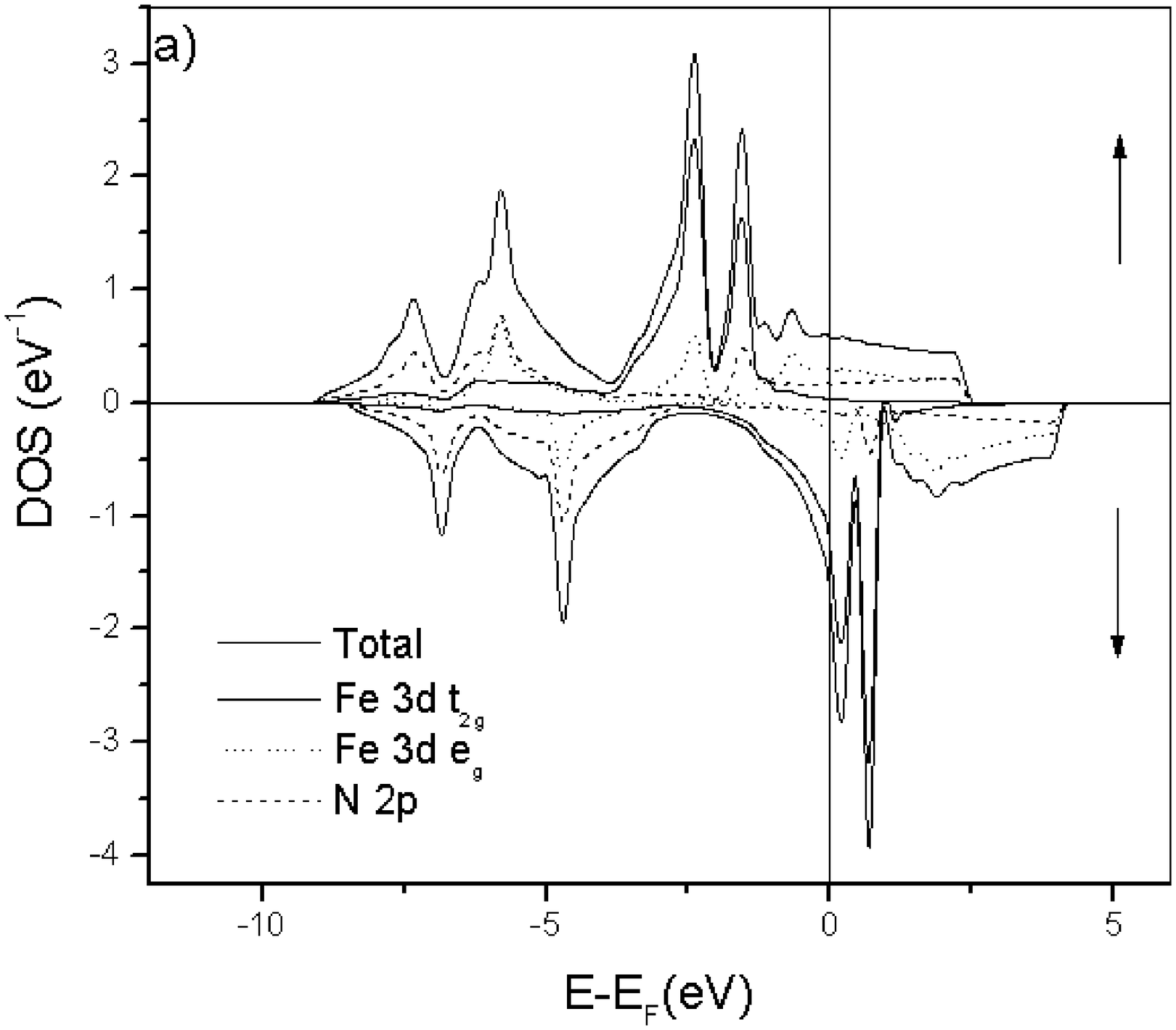}
\includegraphics[width=9.5cm, height=8cm]{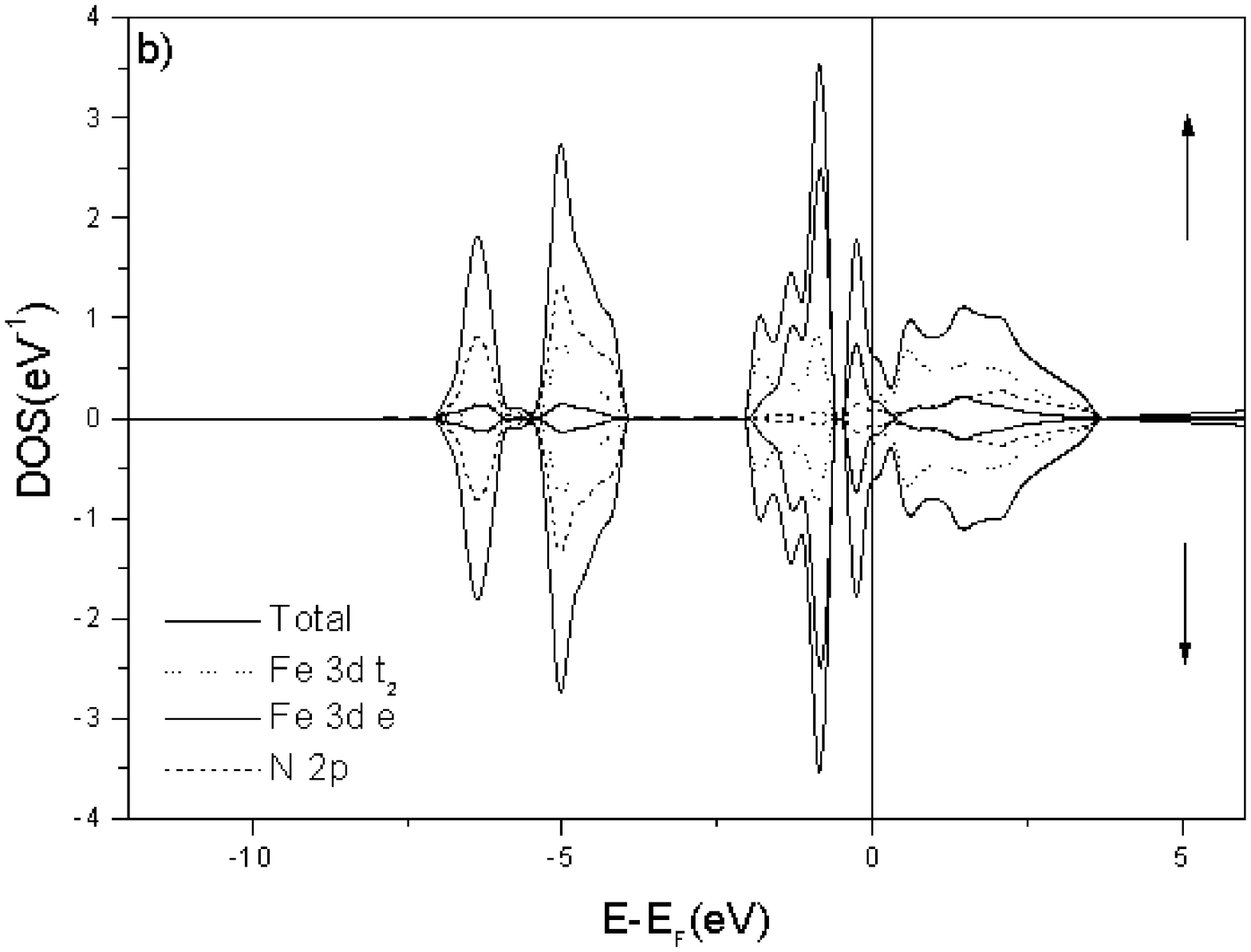}
\caption{Spin polarized total and partial density of states  (PDOS) of a) {\it RS}-FeN and b) {\it ZB}-FeN at equilibrium lattice constants in the ferromagnetic state.}
\end{center}
\end{figure}


\begin{thebibliography}{999}
\bibitem{Jack}
K.~Jack, Proc. R. Soc. London {\bf A11}  34 (1948).
\bibitem{Matar}
S.~F.~Matar, J. Alloys and Compounds {\bf 345} 72 (2002).
\bibitem{ieee}
S.~ F.~ Matar, G. Demazeau and B. Siberchicot. I.E.E.E. Trans. Magn. {\bf 26} 60 (1990). 
\bibitem{Heinman}
N.~Heinman and N.~S.~Kazama, J. Appl. Phys {\bf 52} 3562 (1981).
\bibitem{ouled}
A.~Oueldennaoua, E.~Bauer-Grosse, M.~Foos and C.~Frants, Scr. Metall {\bf 19} 1503 (1985).
\bibitem{Nakag}
H.~Nakagawa, S.~Nasu, M.~Takahashi and F.~Kanamaru, Hyperfine Interact. {\bf 69}  455 (1991).
\bibitem{Hino}
T.~Hinomura and S.~Nasu, Hyperfine Interact. {\bf 111} 221 (1998).
\bibitem{suzuki}
K.~Suzuki, H.~Morita, T.~Kaneko, H.~Yoshida and H.~Fujimori, J. Alloys and Compounds {\bf 201} 11 (1993).
\bibitem{rissanen}
L.~Rissanen, M.~Neubauer, F.P.~Lieb and P.~Schaaf, J. Alloys and Compounds {\bf 274} 74 (1998).
\bibitem{shimizu}
H.~Shimizu, M.~Shirai and N.~Zuzuki, J. Phys. Soc. Jap. {\bf 67}  922 (1998).
\bibitem{filip}
A.~Filippetti and W.E.~Pickett, Phys. Rev. B {\bf 59}, 8397 (1999).
\bibitem{Kong}
Y.~Kong, J. Phys.: Condens. Matter {\bf 12} 4161 (2000).
\bibitem{lukash}
P.~Lukashev and Walter R.~L.~Lambrecht, Phys. Rev. B {\bf 70} 245205 (2004).
\bibitem{eck}
B.~Eck, R.~Dronskowski, M.~Takahashi and S.~Kikkawa, J. Mater. Chem {\bf 9 } 1527 (1999). 
\bibitem{suzuki1}
K.~Suzuki, Y.~Yamaguchi, T.~Kaneko, H.~Yoshida, Y.~Obi, H.~Fujimori and H.~Morita, J. Phys. Soc. Jpn. {\bf 201} 1084 (2001).
\bibitem{kohn}
P.~Honenberg and W.~Kohn, Phys. Rev. {\bf 136}, B864 (1964).
\bibitem{sham}
W.~Kohn and L.J.~Sham, Phys. Rev. {\bf 140},  A1133 (1965).
\bibitem{PBE}
J.P.~Perdew, S.~Burke, and M.~Ernzerhof, Phys. Rev. Lett. {\bf 77} 3865 (1996).
\bibitem{kubler}
A.R.~Williams, J.~K\"ubler and C.D.~Gellat, Phys. Rev. B, {\bf 19} 6094 (1979) .
\bibitem{eyert}
V.~Eyert, Int. J. Quantum Chem. {\bf 77}  1007 (2000).
\bibitem{sgo}
V.\ Eyert and K.-H.\ H\"ock,
Phys.\ Rev.\ B {\bf 57}, 12727 (1998).
\bibitem{stoner}
J.~K\"ubler and V.~Eyert in Electronic and Magnetic Properties of Metal and Ceramics. Buschow K.H.J. Ed (VCH Verlaggesellschaft) (1991).
\bibitem{roy}
D. M. Roy and D. G. Pettifor, J. Phys. F: Met. Phys. {\bf 7} L183-L187  (1977).
\bibitem{friedel}
J.~Friedel, Nuevo Cimento {\bf 10}. 287 (1958).
\bibitem{birch}
G.~Birch, J. Geophys. Res. {\bf 83} 1257 (1978).
\bibitem{mohn}
P.~Mohn and S.~F.~Matar, J. Magn. Magn. Mater {\bf 191} 234 (1999) .
\bibitem{matar1}
S.~F.~Matar and P.~Mohn, Active and Passive Electron. Components {\bf 15} 89 (1993).
\bibitem{janak}
J.~F.~Janak, Phys. Rev. B {\bf 16}  255 (1977).
\bibitem{matar2}
S.~F.~Matar, C. R. Chimie {\bf 5}   539 (2002).
\bibitem{vasp}
VASP code. G. Kresse and J. Hafner, Phys. Rev. B, {\bf 47}, 558 (1993) and G. Kresse and J. Furthm\"uller, Comput. Mat. Sci., {\bf 6}, 15 (1996).
\end{thebibliography}
\end{document}